\documentclass[pra,twocolumn,amsmath]{revtex4-1}
\usepackage{pifont}
\usepackage{graphicx,epsfig,subfigure,dsfont,amssymb,amsmath,amsthm,amsfonts,amsbsy,mathrsfs,amscd}
\usepackage{xcolor}
\usepackage[all]{xy}

\DeclareMathOperator{\tr}{Tr}

\input amssym.def

\begin{document}

\title{Maximum Relative Entropy of Coherence for Quantum Channels}

\smallskip

\author{Zhi-Xiang Jin$^{1,4}$}
\author{Long-Mei Yang$^2$}
\author{Shao-Ming Fei $^{3,4}$}
\thanks{Corresponding author: feishm@cnu.edu.cn}
\author{Xianqing Li-Jost$^4$}
\author{Zhi-Xi Wang$^3$}
\author{Gui-Lu Long$^{2,5,6,7}$}
\author{Cong-Feng Qiao$^{1,8}$}
\thanks{Corresponding author: qiaocf@ucas.ac.cn\\
Z. X. Jin and L. M. Yang contributed equally to this work.}
\affiliation{$^1$School of Physics, University of Chinese Academy of Science, Beijing 100049, China\\
$^2$State Key Laboratory of Low-Dimensional Quantum Physics and Department of Physics, Tsinghua University, Beijing 100084, China\\
$^3$School of Mathematical Sciences, Capital Normal University, Beijing 100048, China\\
$^4$Max-Planck-Institute for Mathematics in the Sciences, 04103 Leipzig, Germany\\
$^5$Frontier Science Center for Quantum Information, Beijing 100084, China\\
$^6$Beijing National Research Center for Information Science and Technology, Beijing 100084, China\\
$^7$ Beijing Academy of Quantum Information Sciences, Beijing 100193, China\\
$^8$ CAS Center for Excellence in Particle Physics, Beijing 100049, China}

\begin{abstract}
Based on the resource theory for quantifying the coherence of quantum channels,
we introduce a new coherence quantifier for quantum channels via maximum relative entropy.
We prove that the maximum relative entropy for coherence of quantum channels is directly related to the
maximally coherent channels under a particular class of superoperations, which results in an
operational interpretation of the maximum relative entropy for coherence of quantum channels.
We also introduce the conception of sub-superchannels and sub-superchannel discrimination.
For any quantum channels, we show that the advantage of quantum channels in sub-superchannel discrimination
can be exactly characterized by the maximum relative entropy of coherence for quantum channels. Similar to the maximum relative entropy of coherence for channels, the robustness of coherence for quantum channels has also been investigated. We show that the maximum relative entropy of coherence for channels provides new operational interpretations of robustness of coherence for quantum channels and illustrates the equivalence of the dephasing-covariant superchannels, incoherent superchannels, and strictly incoherent superchannels in these two operational tasks.

{\bf Keywords:} Maximum Relative Entropy, Quantum Channels, Quantum Coherence, Resource Theory

{\bf PACS:} 03.65.Ta, 03.65.Yz, 03.67.Lx
\end{abstract}

\maketitle

\section{I\MakeLowercase{ntroduction}}  
Quantum coherence characterizes the quantumness of a quantum system, namely, the superposition of a state in a given reference basis. Recently, much attention has been paid to the resource theory of coherence \cite{plenio1,strel1,strel2,strel3,hml,dg,winter,plenio2,gour,vedral}, entanglement \cite{ent}, asymmetry \cite{im,gour1,im1,im2} and non-uniformity \cite{horo, mpp}. As a fundamental feature of quantum physics, quantum coherence plays significant roles in
quantum  metrology \cite{Giovannetti-1,Gyong}, low-temperature thermodynamics \cite{Aberg-1,Gour-1,Horodecki-1,Rudolph-1,Rudolph-2}, nanoscale physics \cite{Plenio-1,Rebentrost-1,Lloyd-1,Li-1,Plenio-2,Levi-1}, and quantum measurement \cite{fh,cna,fsm}.
 The quantification of quantum coherence of quantum states has been extensively investigated, see \cite{lami,tdm,cr,cc} and references therein. In particular, the operational characterization of quantum coherence for quantum states has been presented from the view of resource theory of coherence. It is shown that the max-relative entropy of coherence for any coherent states is directly related to the probability of successful subchannels discrimination \cite{frame-3}.

With respect to the studies on quantum coherence of quantum states,
the coherence of quantum channels has also attracted much attention \cite{Xu-1,Winter-1,Bu-1,Korzekwa-1,Datta-1,Plenio-3,frame-1}.
As most quantum information processes involve and depend on the quantum channels, it is the very act to study the coherence of channels.
Similar to the resource theory of coherence for quantum states, the resource theory of coherence for quantum channels concerns incoherent channels and incoherent superchannels.
However, the general consensus on the set of free superoperations in the resource theory of coherence for quantum channels is still unclear.

In this letter, we introduce a new coherence measure $C_{\max}$ of the coherence for quantum channels based on maximum relative entropy and focus on its operational characterizations.
We prove that the maximum relative entropy for the coherence of a given channel $\phi$ is
achievable with a maximally coherent channel under incoherent superchannels,
which gives rise to an operational interpretation of $C_{\max}$ and shows the equivalence of the incoherent superchannels in an operational task. Besides, we show that the maximum relative entropy of coherence $C_{\max}$ characterizes the sub-superchannel discriminations. In addition,
the maximum relative entropy of coherence for channels also provides new operational interpretations of robustness of coherence for quantum channels and illustrates the equivalence of the dephasing-covariant superchannels, incoherent superchannels and strictly incoherent superchannels in such operational tasks.

Let $H_A$ and $H_B$ be two Hilbert spaces, with dimensions $dim H_A=|A|$ and $dim H_B=|B|$,
and orthonormal bases $\{|j\rangle\}_j$ and $\{|\alpha\rangle\}_{\alpha}$, respectively.
Denote $\mathcal{D}_A$ and $\mathcal{D}_B$ as the sets of all density operators on $H_A$ and $H_B$, respectively, and $\mathcal{C}_{AB}$ the set of all channels from $\mathcal{D}_A$ to $\mathcal{D}_B$.
A channel $\phi\in\mathcal{C}_{AB}$ is an incoherent channel (IC) if
\begin{equation}\label{ic-2}
\Delta^{B}\phi\Delta^{A}=\phi,
\end{equation}
where $\Delta^{A}$ and $\Delta^{B}$ are resource destroying maps \cite{frame-1,Xu-3}.
The set of all $\mathrm{IC}$s is denoted by $\mathcal{IC}_{AB}$.
Similar notations apply to systems $A^\prime$ and $B^\prime$.

A superchannel $\Theta\in\mathcal{SC}_{ABA^{\prime}B^{\prime}}$ is an incoherent superchannel ($\mathrm{ISC}$) if it admits an expression of Kraus operators
$\Theta=\{\mathcal{M}_m\}_m$ such that for each $m$ \cite{frame-1},
\begin{equation}\label{Mm}
\mathcal{M}_m=\sum\limits_{j,\alpha}\mathcal{M}_{mj\alpha }|f(j\alpha)\rangle\langle j\alpha|,
\end{equation}
where $|f(j\alpha)\rangle\in \Big\{{| j'\alpha'\rangle_{j'=1}^{|A'|},}{_{\alpha'=1}^{|B'|}}\Big\}$, and
$\mathcal{SC}_{ABA^{\prime}B^{\prime}}$ is the set of all superchannels $\Theta$ from $\mathcal{C}_{AB}$ to $\mathcal{C}_{A^{\prime}B^{\prime}}$.
The set of all $\mathrm{ISC}$s is denoted by $\mathcal{ISC}_{ABA^{\prime}B^{\prime}}$.

As a framework, any coherence measure $C$ of quantum channels should satisfy \cite{frame-1}:
(i) Faithfulness: $C(\phi)\geqslant0$ for any $\phi\in\mathcal{C}_{AB}$,
  and $C(\phi)=0$ if and only if $\phi\in\mathcal{IC}_{AB}$;
(ii) Nonincreasing under $\mathrm{ISC}_{ABA^{\prime}B^{\prime}}$s: $C(\phi)\geqslant C[\Theta(\phi)]$ for any $\Theta\in\mathcal{ISC}_{ABA^{\prime}B^{\prime}}$;
(iii) Nonincreasing under $\mathrm{ISC}_{ABA^{\prime}B^{\prime}}$s on average: $C(\phi)\geqslant\sum\limits_mp_mC(\phi_m)$
for any $\Theta\in\mathcal{ISC}_{ABA^{\prime}B^{\prime}}$, with $\{\mathcal{M}_m\}_m$ an incoherent expression of $\Theta$, $p_m=\frac{\tr(\mathcal{M}_mJ_\phi\mathcal{M}_m^{\dag})}{|A^{\prime}|}$,
and $J_{\phi_m}=|A^{\prime}|\frac{\mathcal{M}_mJ_{\phi}\mathcal{M}_m^{\dag}}{\tr (\mathcal{M}_mJ_{\phi}\mathcal{M}_m^{\dag})}$;
(iv) Convexity: $C\left(\sum\limits_mp_m\phi_m\right)\leqslant\sum\limits_mp_mC(\phi_m)$
for any $\{\phi_m\}_m\in\mathcal{C}_{AB}$ and probability $\{p_m\}_m$.
Under this framework, the coherence of quantum channels can be calculated via the corresponding Choi matrices. Under this framework the incoherent channels' Choi matrices are diagonal.

For a completely positive trace-preserving (CPTP) operator $\phi\in\mathcal{C}_{AB}$ \cite{Nielsen},
the corresponding Choi matrix is given by \cite{Choi-1,Choi-2}:
$J_{\phi}=\sum\limits_{jk}|j\rangle\langle k|\otimes\phi(|j\rangle\langle k|)=\sum\limits_{jk\alpha\beta}\phi_{jk,\alpha\beta}|j\rangle\langle k|\otimes|\alpha\rangle\langle\beta|$, where $\phi_{jk,\alpha\beta}=\langle\alpha|\phi(|j\rangle\langle k|)|\beta\rangle$ are complex numbers.
It holds that $J_{\phi}\geqslant0$ and $\sum\limits_\alpha\phi_{jk,\alpha\alpha}=\delta_{jk}$.
Note that $\sum\limits_\alpha\phi_{jk,\alpha\alpha}=\delta_{jk}$ is equivalent to $\tr_{B}J_{\phi}=\mathbb{I}_{A}$, with $\mathbb{I}_{A}$ the identity on $H_A$.
In Ref. \cite{frame-1},
the relation between the coherence measures for quantum channels and for quantum states has been presented,
\begin{equation}\label{coherence-1}
C(\phi)=C\left(\frac{J_{\phi}}{|A|}\right), \ \phi\in\mathcal{C}_{AB}.
\end{equation}
A channel is called a channel with maximal coherence if it reaches the maximum for any coherence measure of channels. We also call such channel a maximally coherent channel.

From Eq. \eqref{coherence-1}, one can see that
if a channel $\phi$ is an incoherent channel in the resource theory for the coherence of quantum channels,
$\frac{J_{\phi}}{|A|}$ is an incoherent sate in $H_{A}\otimes H_{B}$.
However, an incoherent state does not always correspond to an incoherent channel,
since if $\phi$ is an incoherent channel, the corresponding Choi matrix $J_{\phi}$ satisfies $\sum\limits_{a}\phi_{jj,\alpha\alpha}=1$ for any $j$,
while for the incoherent state $\sigma\in H_{A}\otimes H_{B}$, $|A|\sigma$ may not be a Choi matrix of some channels.
Namely, the set of the Choi matrices of all quantum channels $\phi\in\mathcal{C}_{AB}$ is a subset of the set of all quantum states in $H_{AB}$.

The corresponding Choi matrix of any superchannel $\Theta\in\mathcal{SC}_{ABA^{\prime}B^{\prime}}$ is given by
$J_{\Theta}=\sum\limits_{jk\alpha\beta}|j\alpha\rangle\langle k\beta|\otimes\Theta(|j\alpha\rangle\langle k\beta|)=\sum\limits_{jk\alpha\beta}\Theta_{jk,\alpha\beta,j^{\prime}k^{\prime},\alpha^{\prime}\beta^{\prime}}
|j\alpha j^{\prime}\alpha^{\prime}\rangle\langle k\beta k^{\prime}\beta^{\prime}|,$
with $\Theta_{jk,\alpha\beta,j^{\prime}k^{\prime},\alpha^{\prime}\beta^{\prime}}=
\langle j^{\prime}\alpha^{\prime}|\Theta(|j\alpha\rangle\langle k\beta|)|k^{\prime}\beta^{\prime}\rangle$.
The superchannel $\Theta$ also has its Kraus operators $\{\mathcal{M}_m\}_m\in\mathcal{SC}_{ABA^{\prime}B^{\prime}}$ such that
\begin{equation}\label{super-1}
J_{\Theta(\phi)}=\sum\limits_m\mathcal{M}_mJ_{\phi}\mathcal{M}_m^{\dag}, \ \forall\phi\in\mathcal{C}_{AB},
\end{equation}
where $\mathcal{M}_m=\sum\limits_{j,\alpha,j',\alpha'}\mathcal{M}_{mj\alpha j'\alpha' }|j'\alpha'\rangle\langle j\alpha|$. From Eq. \eqref{super-1}, one can reasonably assume that
$\sum\limits_m\mathcal{M}_m^{\dag}\mathcal{M}_m=\frac{|A^{\prime}|}{|A|}\mathbb{I}_{AB}$
since $\tr J_{\phi}=|A|$ and $\tr J_{\Theta(\phi)}=|A^{\prime}|$, where $\mathbb{I}_{AB}$ is the identity on $H_{A}\otimes H_{B}$.


We call a superchannel $\Theta\in\mathcal{SC}_{ABA^{\prime}B^{\prime}}$ a dephasing-covariant superchannel (DISC) if $\Theta$ satisfies
$\Delta^{A^{\prime}B^{\prime}}\Theta=\Theta\Delta^{AB}.$
Denote the set of all $\mathrm{DISCs}$ by $\mathcal{DISC}_{ABA^{\prime}B^{\prime}}$.
Let $\{\mathcal{M}_m\}_m$ be the Kraus operators of a superchannel $\Theta\in\mathcal{SC}_{ABA^{\prime}B^{\prime}}$. Equivalently,
a superchannel $\Theta$ is called a $\mathrm{DISC}$ if
$\Delta^{A^{\prime}B^{\prime}}\sum\limits_m\mathcal{M}_m(J_{\phi})\mathcal{M}_m^{\dag}=
\sum\limits_m\mathcal{M}_m\left(\Delta^{AB}(J_{\phi})\right)\mathcal{M}_m^{\dag}$
holds for any $\phi\in\mathcal{C}_{AB}$.


In the resource theory for coherence of quantum channels, from Eq. \eqref{Mm} an incoherent superchannel $\Theta\in\mathcal{SC}_{ABA^{\prime}B^{\prime}}$ has that $\mathcal{M}_m$ transforms the corresponding Choi matrices of incoherent channels to diagonal matrices for each $m$.
A superchannel  $\Theta\in\mathcal{SC}_{ABA^{\prime}B^{\prime}}$ is
called a strictly incoherent superchannel ($\mathrm{SISCs}$) if for each $m$, the Kraus operators $\{\mathcal{M}_m\}_m$ in Eq. (\ref{Mm}) satisfies $\Delta^{A^{\prime}B^{\prime}}\mathcal{M}_m(J_{\phi})\mathcal{M}_m^{\dag}=\mathcal{M}_m\left(\Delta^{AB}(J_{\phi})\right)\mathcal{M}_m^{\dag}$ \cite{GGour}, where $\Delta$ is the free operation defined in Eq. (\ref{ic-2}), i.e., $\langle f( j\alpha)|f(k\beta)\rangle=\delta_{jk,\alpha\beta}$.
We can see that for each $m$, there exists at most one nonzero element in each column (row) of $\mathcal{M}_m$ if $\{\mathcal{M}_m\}_m$ are the Kraus operators of a strictly incoherent superchannel.
Moreover, a strictly incoherent superchannel $\Theta\in\mathcal{SC}_{ABA^{\prime}B^{\prime}}$
can only be defined for the case $|A||B|=|A^{\prime}||B^{\prime}|$.

\section{ M\MakeLowercase{aximum relative entropy of coherence for quantum channels}} 
Given two quantum channels $\phi,\,\widetilde{\phi}\in\mathcal{C}_{AB}$.
The maximum relative entropy of $\phi$ with respect to $\widetilde{\phi}$ is defined by
$D_{\max}\left(\phi|\widetilde{\phi}\right)=\min\left\{\lambda: \, J_{\phi}\leqslant2^{\lambda}J_{\widetilde{\phi}}\right\}.$
We define the measure of coherence, the maximum relative entropy of coherence $C_{\max}$, for a channel $\phi\in\mathcal{C}_{AB}$ by
\begin{equation}\label{max}
C_{\max}(\phi)=\min\limits_{\widetilde{\phi}\in \mathcal{IC}_{AB}}D_{\max}(\phi|\widetilde{\phi}),
\end{equation}
where $\mathcal{IC}_{AB}$ is the set of incoherent channels in $\mathcal{C}_{AB}$.

The maximum relative entropy of coherence $C_{\max}(\phi)$ satisfies the conditions (i), (ii) and (iii) for a well defined measure of coherence for quantum channels.
Obviously $C_{\max}(\phi)\geqslant0$, and $D_{\max}(\phi|\widetilde{\phi})=0$
if and only if $J_\phi=J_{\widetilde{\phi}}$ \cite{max-1}. Hence, $C_{\max}(\phi)=0$ if and only if $\phi\in\mathcal{IC}_{AB}$.
For the proof that $C_{\max}(\phi)$ satisfies the conditions (ii) and (iii), see Section A in Supplemental Material.
Thus, the maximum relative entropy of coherence $C_{\max}$ is a bona fide measure of coherence of quantum channels.

$C_{\max}$ may be not convex since $D_{\max}$ is not jointly convex. However, for $\phi=\sum_mp_m\phi_m$, we have $C_{\max}(\phi)\leq\max_mC_{\max}(\phi_m)$.
Suppose $\phi_m^*$ is the optimal channel such that $C_{\max}(\phi_m)=D_{\max}(\phi_m|\phi_m^*)$. From the result that $D_{\max}(\sum_mp_m{\phi_m}|\sum_mp_m{\widetilde{\phi}_m})\leq \max_mD_{\max}({\phi_m}|{\widetilde{\phi}_m})$ \cite{max-2}, we have $C_{\max}(\phi)\leq D_{\max}(\sum_mp_m{\phi_m}|\sum_mp_m{\phi_m^*})\leq \max_mD_{\max}({\phi_m}|{\phi_m^*})=\max_mC_{\max}(\phi_m)$. Moreover, define by convex roof extension of $C_{\max}$,
$$
{\widetilde C}_{\max}(\phi)=\min\sum_i\lambda_i C_{\max}(\psi_i),
$$
where $\sum_i\lambda_i=1$, the minimum is taken over all the pure quantum channel decompositions of channel $\phi\in\mathcal{C}_{AB}$, $\phi=\sum_i\lambda_i\psi_i$, $\psi_i\in\mathcal{C}_{AB}$ are pure quantum channels such that $\frac{J_{\psi_i}}{|A|}$ are pure states.
It can be shown that ${\widetilde C}_{\max}$ is a proper coherence measure of quantum channels with the property of convexity, see Section B in the Supplemental Material. We show next that $C_{\max}(\phi)$ has direct operational meanings.

\section{M\MakeLowercase{aximum overlap with maximally coherent channels}}
At first we show that $C_{\max}$ characterizes the maximum overlap with a maximally coherent channel that can be achieved by $\mathcal{ISC}_{ABA^{\prime}B^{\prime}}$, see section C in the Supplemental Material.

{\it Theorem 1.} -- Given a quantum channel $\phi\in\mathcal{C}_{AB}$, we have
\begin{equation}\label{max-4}
2^{C_{\max}(\phi)}=\frac{|B'|}{|A'|}\max\limits_{\Theta\in\mathcal{ISC}_{ABA^{\prime}B^{\prime}}}
F\left(\Theta(\phi),\Phi\right)^2,
\end{equation}
where $F(\phi,\widetilde{\phi}):=\tr\Big[\sqrt{J_\phi}\sqrt{ J_{\widetilde{\phi}}}\Big]$ is the fidelity between the two quantum channels,
$\Phi$ is a maximally coherent channel in $\mathcal{C}_{A^{\prime}B^{\prime}}$, $|A||B|\leqslant|A^{\prime}||B^{\prime}|$.

In fact, Theorem 1 holds also for $\mathcal{DISC}_{ABA^{\prime}B^{\prime}}$ and $\mathcal{SISC}_{ABA^{\prime}B^{\prime}}$, see section D in the Supplemental Material.

{\it Remark} -- Although ISC, DISC and SISC are different types of free superchannels in the resource theory of coherence for quantum channels, they have the same behavior in the maximum overlap with the maximally coherent channels. Specially, for $|A|=|A^\prime|=1$, a coherence measure for quantum channels degenerates to a coherence measure for quantum states. Then from Eq. (\ref{max-4}), given a quantum state $\rho\in\mathcal{D}_{B}$ and $\phi\in\mathcal{C}_{BB'}$, one gets  $2^{C_{\max}(\rho)}=|B'|\max\limits_{\phi} F[\phi(\rho),|\Psi\rangle\langle\Psi|]^2,$
where $F(\rho,\sigma)=\tr [\sqrt{\rho}\sqrt{\sigma}]$ is the fidelity between states $\rho$ and $\sigma$,
$|\Psi\rangle$ is an arbitrary maximally coherent state in $\mathcal{D}_{B'}$,
and $\phi$ belongs to either incoherent operation (IO), dephasing covariant operation (DIO), strictly incoherent operation (SIO), which
are the set of incoherent operations, dephasing-covariant operations and strictly incoherent operations
in the resource theory for the coherence of quantum states, respectively.

\section{M\MakeLowercase{aximum advantage achievable in sub-superchannel discrimination}} 
The subchannel and subchannel discrimination has been introduced in Ref. \cite{subchannel}. Similar to the construction of subchannel discrimination, we introduce the conception of sub-superchannels and sub-superchannel discrimination as follows.
A linear completely positive $\Theta\in\mathcal{SC}_{ABA^{\prime}B^{\prime}}$ with its Kraus operators $\{\mathcal{M}_m\}_m$ satisfying $\sum\limits_m\mathcal{M}_m^{\dag}\mathcal{M}_m\leqslant\frac{|A^{\prime}|}{|A|}\mathbb{I}_{AB}$ is called a sub-superchannel.
If the Kraus operators $\{\mathcal{M}_m\}_m$ satisfy
$\sum\limits_m\mathcal{M}_m^{\dag}\mathcal{M}_m=\frac{|A^{\prime}|}{|A|}\mathbb{I}_{AB}$,
then $\Theta$ is called a superchannel.
An instrument $\mathfrak{J}=\{\Theta_k\}_k$ for a superchannel $\Theta$ is a collection of sub-superchannels $\Theta_k$
with $\Theta=\sum\limits_k\Theta_k$.

Give an instrument $\mathfrak{J}=\{\Theta_k\}_k$ for a superchannel $\Theta\in\mathcal{SC}_{ABA^{\prime}B^{\prime}}$.
Consider the positive operator-valued measure (POVM) $\{M_{k^{\prime}}\}_{k^{\prime}}$ with
$\sum\limits_{k^{\prime}}M_{k^{\prime}}=\mathbb{I}_{A^{\prime}B^{\prime}}$.
The joint probability of $\Theta_k$ and $M_{k^{\prime}}$ for a channel $\phi$ is
$p(k,k^{\prime})=\frac{\tr\left[M_{k^{\prime}}\left(\sum\limits_m\mathcal{M}_{mk}J_{\phi}\mathcal{M}_{mk}^{\dag}\right)\right]}{|A^{\prime}|}
=p(k^{\prime}|k)p(k)$,
where $p(k)=\frac{\tr\left[\left(\sum\limits_m\mathcal{M}_{mk}J_{\phi}\mathcal{M}_{mk}^{\dag}\right)\right]}{|A^{\prime}|}$
is the probability of the sub-superchannel $\Theta_k$ for $\phi$,
$p(k^{\prime}|k)$  is the conditional probability of the outcome $k^{\prime}$ given that the sub-superchannel $\Theta_k$ takes place,
and $\Theta_k=\{\mathcal{M}_{mk}\}_m$ are Kraus operators of $\Theta_k$.
Then the probability of successfully discriminating the sub-superchannels in instrument $\mathfrak{J}$
by POVM $\{M_{k^{\prime}}\}_{k^{\prime}}$ is given by
\begin{eqnarray*}
p_{succ}(\mathfrak{J},\{M_{k^{\prime}}\}_{k^{\prime}},\phi)=
\frac{\sum\limits_k\tr\left[M_{k^{\prime}}\left(\sum\limits_m\mathcal{M}_{mk}J_{\phi}\mathcal{M}_{mk}^{\dag}\right)\right]}{|A^{\prime}|}
\end{eqnarray*}
for $\phi\in\mathcal{C}_{AB}$.

The optimal probability of success in sub-superchannel discrimination of $\mathfrak{J}$
over all POVMs is given by
\begin{eqnarray*}
p_{succ}(\mathfrak{J},\phi)=\max\limits_{\{M_{k^{\prime}}\}_{k^{\prime}}}p_{succ}(\mathfrak{J},\{M_{k^{\prime}}\}_{k^{\prime}},\phi).
\end{eqnarray*}
If we restrict $\phi$ to incoherent channels, the optimal probability of success among all incoherent channels is given by
\begin{eqnarray*}
p_{succ}^{ISCO}(\mathfrak{J})=\max\limits_{\widetilde{\phi}\in\mathcal{IC}_{AB}}p_{succ}(\mathfrak{J},\widetilde{\phi}).
\end{eqnarray*}
Then we have the following theorem, see proof in section E in the Supplemental Material.

{\it Theorem 2.} --For any quantum channel $\phi\in\mathcal{C}_{AB}$, $2^{C_{\max}(\phi)}$ is the maximal advantage achievable by $\phi$
compared with all incoherent channels in sub-superchannel discriminations of incoherent superchannel instruments:
\begin{equation*}
2^{C_{\max}(\phi)}=\max\limits_{\mathfrak{J}}\frac{p_{succ}(\mathfrak{J},\phi)}{p_{succ}^{ISCO}(\mathfrak{J})},
\end{equation*}
where $\mathfrak{J}$ denotes the incoherent instrument of superchannel $\Theta\in\mathcal{ISC}_{ABA^{\prime}B^{\prime}}$, $|A||B|\leqslant|A^{\prime}||B^{\prime}|$.

Theorem 2 also holds for $\Theta\in\mathcal{DISC}_{ABA^{\prime}B^{\prime}}$ and
$\Theta\in\mathcal{SISC}_{ABA^{\prime}B^{\prime}}$.
Theorem 2 shows that the advantage of quantum channels in sub-superchannel discriminations
can be exactly captured by $C_{\max}$, and provides another operational interpretation of $C_{\max}$.
One should note here that although $\tilde {C}_{max}$ is a proper coherence measure of quantum channels by the convex roof extension of $C_{max}$, it can not characterize the advantage of quantum channels in sub-superchannel discrimination task, since the optimal incoherent instrument $\mathfrak{J}_i$ of $\psi_i$ for the optimal pure quantum channel decomposition of $\phi=\sum_i\lambda_i \psi_i$ may be different and $2^{C_{\max}}$ in Theorem 2 is not linear.

\section{ $C_{\max}$\MakeLowercase{ and robustness of coherence measure for quantum channels}} 
Similar to the robustness of coherence (ROC) of quantum states \cite {cna},
we can define the ROC of a quantum channel $\phi\in\mathcal{C}_{AB}$ as
\begin{equation}\label{roc}
C_R(\phi)=\min_{\widetilde{\phi}\in\mathcal{C}_{AB}}\bigg\{s\geq0\bigg|\frac{J_{\phi}+sJ_{\widetilde\phi}}{1+s}=J_{\phi^\prime}, \phi^\prime\in \mathcal{IC}_{AB}\bigg\},
\end{equation}
namely, the minimum weight of another channel $\widetilde{\phi}$ such that its convex mixture with $\phi$ yields an incoherent channel $\phi^\prime$.
We now prove that the ROC is a well defined measure of coherence for channels. By definition $C_R(\phi)\geq0$ and $C_R(\phi)=0$ if and only if $\phi\in\mathcal{IC}_{AB}$.
For the fact that $C_R$ satisfies the conditions (iii) and (iv), see Section F in the Supplemental Material. Notice that conditions (iii) and (iv) automatically imply condition (ii). This can be seen as follows:
$C_R[\Theta(\phi)]=C_R\left(\sum\limits_mp_m\phi_m\right)\leq \sum\limits_mp_mC_R(\phi_m)\leq C_R(\phi)$,
where the first inequality is due to the convexity of $C_R$, and the second inequality is due to the monotonicity under selective measurements on average. Thus, the robustness of coherence $C_R$ is a bona fide measure of coherence of quantum channels.

It can be shown that, see Section F in the Supplemental Material,
\begin{eqnarray}\label{lroc}
C_R(\phi)=\min_{\phi^\prime\in \mathcal{IC}_{AB}}\{s\geq0|J_\phi\leq (1+s)J_{\phi^\prime}\}.
\end{eqnarray}
Therefore, we have $2^{C_{\max}(\phi)}=1+C_R(\phi)$. The operational interpretations of $C_{\max}$ on the maximum overlap with maximally coherent channels and the subsuperchannel discriminations also give rise to the operational interpretations of $C_R$.
The ROC of quantum states plays an important role in phase channel discriminations \cite{cna}, which corresponds in particular to the subsuperchannel discrimination in dephasing-covariant instruments for $|A|=|A^\prime|=1$. As for a closed form between $C_{\max}$ and $C_R$, let us consider a pure quantum channel $\psi$ and its Choi matrix $J_\psi=|A||\psi\rangle\langle \psi|$ with $|\psi\rangle=\sum_{j,k}^{|A|,|B|}\lambda_{jk}|j\rangle |k\rangle$ and $\sum_k|\lambda_{jk}|^2=\frac{1}{|A|}$. Then $C_{\max}(\psi)=2\log(\sum_{j,k}|\lambda_{jk}|)^2$. In particular, for the maximally coherent channel $\Psi$ with the Choi matrix $J_\Psi=|A||\Psi\rangle\langle \Psi|$ with $|\Psi\rangle=\frac{1}{\sqrt{|A||B|}}\sum_{j,k}^{|A|,|B|}e^{i\theta_{jk}}|j\rangle |k\rangle$, we have $C_{\max}(\Psi)=\log|A||B|$, which is the maximum value of $C_{\max}$ for channels $\psi\in\mathcal{C}_{AB}$.
Above all, the two coherence measures $C_{\max}(\phi)$ and $C_R(\phi)$ of quantum channels are ordered such that
$C_{\max}(\phi_1)\leqslant C_{\max}(\phi_2)$ if and only if $C_R(\phi_1)\leqslant C_R(\phi_2)$.

\section{C\MakeLowercase{onclusion}} 
We have presented a bona fide measure of coherence for quantum channels, the maximum relative entropy of coherence $C_{\max}$, which has explicit operational interpretations. It characterizes the maximum overlap with the maximally coherent channels under DISC, ISC and SISC, as well as
the maximum advantage achievable by coherent quantum channels compared with incoherent quantum channels in sub-superchannel discriminations for all dephasing-covariant superchannel instruments, incoherent superchannel instruments and strictly incoherent superchannel instruments. $C_{\max}$ also provides new operational interpretations of robustness of coherence $C_R$ for quantum channels and illustrates the equivalence of DISC, ISC and SISC under these two operational tasks. It has been found that $C_{\max}$ and $C_R$ have the same ordering for quantum channels. In addition, a coherence measure for quantum channels degenerates to a coherence measure for quantum states for $|A|=|A^\prime|=1$.
These results may highlight the understanding of the operational resource theory of coherence for quantum channels.

\bigskip
\noindent {\bf Acknowledgments}
This work is supported by the NSF of China under Grant Nos. 11675113, 11847209, 61727801 and 12075159,
the China Postdoctoral Science Foundation funded project No. 2019M650811 and the China Scholarship Council No. 201904910005,
Shenzhen Institute for Quantum Science and Engineering, Southern University of Science and Technology (Grant No. SIQSE202005),
the Key Project of Beijing Municipal Commission of Education (Grant No. KZ201810028042), Beijing Natural Science Foundation (Z190005), the Academician Innovation Platform of Hainan Province, and Academy for Multidisciplinary Studies, Capital Normal University, the China Postdoctoral Science Foundation funded project No. 2019M650811 and the China Scholarship Council No. 201904910005.

\end{document}


\title{Maximum Relative Entropy for Coherence of Quantum Channels}

\smallskip

\author{Zhi-Xiang Jin$^{1,4}$}
\author{Long-Mei Yang$^2$}
\author{Shao-Ming Fei $^{3,4}$}
\email{feishm@cnu.edu.cn }
\author{Xianqing Li-Jost$^4$}
\author{Zhi-Xi Wang$^3$}
\author{Gui-Lu Long$^{2,5,6,7}$}
\author{Cong-Feng Qiao$^{1,8}$}
\email{qiaocf@ucas.ac.cn}
\affiliation{$^1$School of Physics, University of Chinese Academy of Science, Beijing 100049, China\\
$^2$State Key Laboratory of Low-Dimensional Quantum Physics and Department of Physics, Tsinghua University, Beijing 100084, China\\
$^3$School of Mathematical Sciences, Capital Normal University, Beijing 100048, China\\
$^4$Max-Planck-Institute for Mathematics in the Sciences, 04103 Leipzig, Germany\\
$^5$Frontier Science Center for Quantum Information, Beijing 100084, China\\
$^6$Beijing National Research Center for Information Science and Technology, Beijing 100084, China\\
$^7$ Beijing Academy of Quantum Information Sciences, Beijing 100193, China\\
$^8$ CAS Center for Excellence in Particle Physics, Beijing 100049, China}

\maketitle

\section*{A: Properties of $C_{\max}$}

\subsection*{ii):  Nonincreasing under $\mathrm{ISC}_{ABA^{\prime}B^{\prime}}$s}

Note that $D_{\max}$ is monotone under completely positive and trace
preserving (CPTP) maps \cite{c-1}.
Denote $\Theta^{\prime}=\frac{|A|}{|A^{\prime}|}\Theta$ with $\Theta\in\mathrm{ISC}_{ABA^{\prime}B^{\prime}}$.
Thus, $\Theta^{\prime}$ is a CPTP map and $J_{\Theta{\prime}(\phi)}=\frac{|A|}{|A^{\prime}|}J_{\phi}$ correspondingly.
Then
\begin{equation}\label{rl}
\begin{array}{rl}
D_{\max}\left(\Theta(\phi)|\Theta(\widetilde{\phi})\right)
&=\min\left\{\lambda|J_{\Theta(\phi)}\leqslant2^{\lambda}J_{\Theta(\widetilde{\phi})}\right\}\\[2.00mm]
&=\min\left\{\lambda|J_{\Theta^{\prime}(\phi)}\leqslant2^{\lambda}J_{\Theta^{\prime}(\widetilde{\phi})}\right\}\\[2.00mm]
&\leqslant \min\left\{\lambda|J_{\phi}\leqslant2^{\lambda}J_{\widetilde{\phi}}\right\}\\[2.00mm]
&=D_{\max}(\phi|\widetilde{\phi}).
\end{array}
\end{equation}
Therefore,
\begin{equation*}
\begin{array}{rl}
C_{\max}\left(\Theta(\phi)\right)
&=\min\limits_{\widetilde{\phi}\in\mathcal{IC}_{AB}} D_{\max}\left(\Theta(\phi)|\widetilde{\phi}\right)\\[2.00mm]
&\leqslant\min\limits_{\widetilde{\phi}\in\mathcal{IC}_{AB}} D_{\max}\left(\Theta(\phi)|\Theta(\widetilde{\phi})\right)\\[2.00mm]
&\leqslant \min\limits_{\widetilde{\phi}\in\mathcal{IC}_{AB}} D_{\max}\left(\phi|\widetilde{\phi}\right)\\[2.00mm]
&=C_{\max}(\phi),
\end{array}
\end{equation*}
where the second line follows from that $\Theta(\widetilde{\phi})\in\mathcal{IC}_{AB}$, and the third line follows from (\ref{rl}).

\subsection*{iii): Nonincreasing under $\mathrm{ISC}_{ABA^{\prime}B^{\prime}}$s on average}

Assume that $\widetilde{\phi}^{*}\in\mathcal{IC}_{AB}$ be an optimal incoherent channel
such that $C_{\max}(\phi)=D_{\max}(\phi|\widetilde{\phi}^{*})$. Due to one to one correspondence between $\phi$ and $J_\phi$, we denote  $D_{\max}\left(J_{\phi}|J_{\widetilde{\phi}}\right)=D_{\max}\left(\phi|\widetilde{\phi}\right)$ for convenience.
Let $J_{\widetilde{\phi}_m^*}=|A^{\prime}|\frac{\mathcal{M}_mJ_{\widetilde{\phi}^*}\mathcal{M}_m^{\dag}}{\tr (\mathcal{M}_mJ_{\widetilde{\phi}^*}\mathcal{M}_m^{\dag})}$.
We have
\begin{equation*}
\begin{array}{rl}
&\sum\limits_m p_m C_{\max}(\phi_m)\\[2.00mm]
&\ \ = \min\limits_{\widetilde{\phi}_m\in\mathcal{IC}_{AB}}
\sum\limits_m p_m D_{\max}\left(\phi_m|\widetilde{\phi}_m\right)\\[2.00mm]
& \ \ \leqslant\sum\limits_m p_m D_{\max}\left(J_{\phi_m}|J_{\widetilde{\phi}^*_m}\right)\\[2.00mm]
& \ \ \leqslant\sum\limits_m D_{\max}\left(\mathcal{M}_mJ_{\phi}\mathcal{M}_m^{\dag}|\mathcal{M}_mJ_{\widetilde{\phi}^*}\mathcal{M}_m^{\dag}\right)\\[2.00mm]
& \ \ \leqslant\sum\limits_m D_{\max}(\tr_E[\mathbb{I}\otimes|m\rangle\langle m|UJ_{\phi}\otimes|\alpha\rangle\langle\alpha|
U^{\dag}\mathbb{I}\otimes|m\rangle\langle m|]|\\[2.00mm]
& \ \ \ \ \ \ \ \ \ \ \tr_E[\mathbb{I}\otimes|m\rangle\langle m|UJ_{\widetilde{\phi}^*}\otimes|\alpha\rangle\langle\alpha|
U^{\dag}\mathbb{I}\otimes|m\rangle\langle m|])\\[2.00mm]
& \ \ \leqslant\sum\limits_m D_{\max}([\mathbb{I}\otimes|m\rangle\langle m|UJ_{\phi}\otimes|\alpha\rangle\langle\alpha|
U^{\dag}\mathbb{I}\otimes|m\rangle\langle m|]|\\[2.00mm]
& \ \ \ \ \ \ \ \ \ \ [\mathbb{I}\otimes|m\rangle\langle m|UJ_{\widetilde{\phi}^*}\otimes|\alpha\rangle\langle\alpha|
U^{\dag}\mathbb{I}\otimes|m\rangle\langle m|])\\[2.00mm]
& \ \ =D_{\max}(UJ_{\phi}\otimes|\alpha\rangle\langle\alpha|U^{\dag}|UJ_{\widetilde{\phi}^*}\otimes|\alpha\rangle\langle\alpha|U^{\dag})\\[2.00mm]
& \ \ =D_{\max}(J_{\phi}\otimes|\alpha\rangle\langle\alpha||J_{\widetilde{\phi}^*}\otimes|\alpha\rangle\langle\alpha|)\\[2.00mm]
& \ \ =D_{\max}(J_{\phi}|J_{\widetilde{\phi}^*})\\[2.00mm]
& \ \ =C_{\max}(\phi),
\end{array}
\end{equation*}
where the second inequality is due to the proof of Theorem 1 in \cite{c-1},
the third inequality is due to the fact that there exist an extended Hilbert space $H_E$,
a pure $|\alpha\rangle\in H_E$ and a global unitary $U$ on $H_{AB}\otimes H_E$ such that
$\tr_E[\mathbb{I}\otimes|m\rangle\langle m|UJ_{\phi}\otimes|\alpha\rangle\langle\alpha|
U^{\dag}\mathbb{I}\otimes|m\rangle\langle m|]=\frac{\mathcal{M}_mJ_{\phi}\mathcal{M}_{m}^{\dag}}{|A^{\prime}|}$ \cite{c-2},
the forth inequality is due to the fact that $D_{\max}$ is monotone under partial trace \cite{c-1},
and the second equality is due to the fact that for any set of mutually orthogonal projectors $\{P_n\}$,
$D_{\max}\left(\sum\limits_nP_nJ_{\phi_1}P_n|P_nJ_{\phi_2}P_n\right)=\sum\limits_nD_{\max}\left(P_nJ_{\phi_1}P_n|P_nJ_{\phi_2}P_n\right)$ \cite{c-1}.

\section*{B: Coherence measure with convexity induced from $C_{\max}$}

Due to the definition of ${\widetilde C}_{\max}$ and the properties of $C_{\max}$, the positivity and convexity of ${\widetilde C}_{\max}$ are obvious. 
We prove that ${\widetilde C}_{\max}$ is nonincreasing on average under $\mathrm{ISC}_{ABA^{\prime}B^{\prime}}$s.

Let $\phi=\sum_i\lambda_i\psi_i$ be the optimal pure quantum channel decomposition of $\phi$ such that ${\widetilde C}_{\max}(\phi)=\sum_i\lambda_i C_{\max}(\psi_i)$. Denote $\Theta^{\prime}=\frac{|A|}{|A^{\prime}|}\Theta$ for any incoherent superchannel $\Theta=\{\mathcal{M}_m\}_m\in \mathrm{ISC}_{ABA^{\prime}B^{\prime}}$ with $\mathcal{M}_m=\sum\limits_{j\alpha}\mathcal{M}_{mj\alpha}|f(j\alpha)\rangle\langle j\alpha|$.
Then $\Theta^{\prime}=\{\mathcal{M^\prime}_m\}_m=\{\sqrt{\frac{|A|}{|A^\prime}}\mathcal{M}_m\}_m$ is a CPTP map. We have
\begin{eqnarray*}
J_{\phi_m}&&=\frac{\mathcal{M}_mJ_{\phi}\mathcal{M}_m^{\dag}}{p_m}\nonumber\\
&&=\sum_i \frac{\lambda_i}{p_m}\mathcal{M}_mJ_{\psi_i}\mathcal{M}_m^{\dag}\nonumber\\
&&=\sum_i \frac{\lambda_iq^{(m)}_i}{p_m}J_{\phi^{(m)}_i},
\end{eqnarray*}
where $p_m=\frac{\tr(\mathcal{M}_mJ_\phi\mathcal{M}_m^\dag)}{|A'|}$, and $q^{(m)}_i=\frac{\tr(\mathcal{M}_mJ_{\psi_i}\mathcal{M}_m^{\dag})}{|A'|}$. 
Thus we have 
\begin{eqnarray}\label{b1}
{\widetilde C}_{\max}(\phi_m)\leq \sum_i\frac{\lambda_iq^{(m)}_i}{p_m} C_{\max}(\phi^{(m)}_i).
\end{eqnarray}
Then we have the following
\begin{eqnarray*}
&&\sum_mp_m{\widetilde C}_{\max}(\phi_m)\nonumber\\
&&\leq \sum_{i,m}\lambda_iq^{(m)}_iC_{\max}(\phi^{(m)}_i)\nonumber\\
&&= \sum_{i,m}\lambda_iq^{(m)}_iC_{\max}\left(\frac{J_{\phi^{(m)}_i}}{|A^{\prime}|}\right)\nonumber\\
&&=\sum_{i}\lambda_i\sum_{m}q^{(m)}_i\log\left(1+C_{l_1}\left(\frac{J_{\phi^{(m)}_i}}{|A^{\prime}|}\right)\right)\nonumber\\
&&\leq \sum_{i}\lambda_i\log\left(1+\sum_{m}q^{(m)}_iC_{l_1}\left(\frac{J_{\phi^{(m)}_i}}{|A^{\prime}|}\right)\right)\nonumber\\
&&\leq\sum_{i}\lambda_i\log\left(1+C_{l_1}\left(\sum_{m}\mathcal{M^\prime}_m\frac{J_{\psi_i}}{|A|}\mathcal{M^\prime}_m^{\dag}\right)\right)\nonumber\\
&&\leq \sum_{i}\lambda_i\log\left(1+C_{l_1}\left(\frac{J_{\psi_i}}{|A|}\right)\right)\nonumber\\
&&=\sum_{i}\lambda_iC_{\max}\left(\frac{J_{\psi_i}}{|A|}\right)\nonumber\\
&&={\widetilde C}_{\max}(\phi),
\end{eqnarray*}
where the first inequality comes from Eq. (\ref{b1}); one gets the first equality from the Eq. (3) in the main text; the second equality comes from the fact that for any pure state $\rho=|\psi\rangle\langle \psi|$, $C_{\max}(\rho)=\log(1+C_{l_1}(\rho))$ \cite{cna} with $C_{l_1}(\rho)=\sum_{i\ne j}|\langle i|\rho |j\rangle|$ the $l_1$-norm coherence of the state $\rho$; the second inequality comes from the concavity of logarithm; using $\sum_m\mathcal{M^\prime}_m\left(\frac{J_{\psi_i}}{|A|}\right)\mathcal{M^\prime}_m^{\dag}=\sum_m q^{(m)}_i\frac{J_{\phi^{(m)}_i}}{|A^{\prime}|}$ with $\{\mathcal{M^\prime}_m\}_m=\{\sqrt{\frac{|A|}{|A^\prime}}\mathcal{M}_m\}_m$ a incoherent CPTP map, we get the third inequality; the fourth inequality due to the monotonicity of $C_{l_1}$ under the incoherent operations.

\section*{C: Proof of Theorem 1}

{\it Restatement of Theorem 1.} -- Given a quantum channel $\phi\in\mathcal{C}_{AB}$, we have
\begin{equation}\label{max-4}
2^{C_{\max}(\phi)}=\frac{|B'|}{|A'|}\max\limits_{\Theta\in\mathcal{ISC}_{ABA^{\prime}B^{\prime}}}
F\left(\Theta(\phi),\Phi\right)^2,
\end{equation}
where $F(\phi,\widetilde{\phi}):=\tr\Big[\sqrt{J_\phi}\sqrt{ J_{\widetilde{\phi}}}\Big]$ is the fidelity between the two quantum channels,
$\Phi$ is a maximally coherent channel in $\mathcal{C}_{A^{\prime}B^{\prime}}$, $|A||B|\leqslant|A^{\prime}||B^{\prime}|$.

{\sf Proof.} An outline of the proof of Theorem 1 is as follows.
First of all, let's consider the right side of Eq. (\ref{max-4}), we have
\begin{eqnarray}\label {pfth}
\frac{|B'|}{|A'|}F(\Theta(\phi),\Phi)^2
&&=\frac{|B|}{|A|}\tr\left[J_{\Theta(\phi)}\left(\frac{|A||B^{\prime}|}{|A^{\prime}||B|}J_{\Phi}\right)\right]\nonumber\\
&&=\frac{|B|}{|A|}\tr\left[J_{\phi}\left(J_{\frac{|A||B^{\prime}|}{|A^{\prime}||B|}\Theta^{\dag}\left(\Phi\right)}\right)\right]\nonumber\\
&&=\tr\left[\frac{J_{\phi}}{|A|}(|B|J_{\phi^{\prime}})\right],
\end{eqnarray}
where $\phi^{\prime}=\frac{|A||B^{\prime}|}{|A^{\prime}||B|}\Theta^{\dag}\left(\Phi\right)$ is a channel in $\mathcal{C}_{AB}$. 

Our task now is to prove the one to one correspondence between the set of the incoherent superchannel $\{\Theta|\Theta\in\mathcal{ISC}_{ABA^{\prime}B^{\prime}}\}$ and the set of the channel $\{\phi^{\prime}|J_{\phi^{\prime}}\geqslant0, \Delta(|B|J_{\phi^{\prime}})=\mathbb{I}_{AB}, \phi^{\prime}\in\mathcal{C}_{AB}\}$.

On the one hand, for any $\Theta\in\mathcal{ISC}_{ABA^{\prime}B^{\prime}}$, by the definition of the incoherent superchannel,  there exists a set of Kraus operators $\{\mathcal{M}_m\}_m$ such that for each $m$,
\begin{equation*}
\mathcal{M}_m=\sum\limits_{j\alpha}\mathcal{M}_{mj\alpha}|f(j\alpha)\rangle\langle j\alpha|,
\end{equation*}
where $|f(j\alpha)\rangle\in \Big\{{| j'\alpha'\rangle_{j'=1}^{|A'|},}{_{\alpha'=1}^{|B'|}}\Big\}$.
Then we have
\begin{equation*}\label{Mm-1}
\begin{array}{rl}
&\sum\limits_{m}\langle j\alpha|\mathcal{M}_m^{\dag}(J_{\Phi})\mathcal{M}_m|j\alpha\rangle\\
& \ \ =\frac{1}{|B^{\prime}|}\sum\limits_{\substack{{m,j^{\prime},\alpha^{\prime}}\\ {k^{\prime},\beta^{\prime}}}}
e^{i(\theta_{j^{\prime}\alpha^{\prime}}-\theta_{k^{\prime}\beta^{\prime}})}
\langle j\alpha|\mathcal{M}_m^{\dag}|j^{\prime}\alpha^{\prime}\rangle\langle k^{\prime}\beta^{\prime}|\mathcal{M}_m|j\alpha\rangle\\
& \ \ =\frac{1}{|B^{\prime}|}\sum\limits_{mj^{\prime}\alpha^{\prime}}
\langle j\alpha|\mathcal{M}_m^{\dag}|j^{\prime}\alpha^{\prime}\rangle\langle j^{\prime}\alpha^{\prime}|\mathcal{M}_m|j\alpha\rangle\\
& \ \ =\frac{1}{|B^{\prime}|}\sum\limits_{m}\langle j\alpha|\mathcal{M}_m^{\dag}\mathcal{M}_m|j\alpha\rangle\\
& \ \ =\frac{|A^{\prime}|}{|A||B^{\prime}|},
\end{array}
\end{equation*}
where the first equation is because for any maximally coherent channel $\Phi\in\mathcal{C}_{A^{\prime}B^{\prime}}$,
there exists a set of $\{\theta_{j^{\prime}\alpha^{\prime}}\}_{j^{\prime}=1}^{|A^{\prime}|},_{\alpha^{\prime}=1}^{|B^{\prime}|}$
such that
$\frac{J_{\Phi}}{|A^{\prime}|}=|\Phi\rangle\langle\Phi|$,
with $|\Phi\rangle=\frac{1}{\sqrt{|A^{\prime}||B^{\prime}|}}\sum\limits_{j^{\prime}=1}^{|A^{\prime}|}
\sum\limits_{\alpha^{\prime}=1}^{|B^{\prime}|}e^{i\theta_{j^{\prime},\alpha^{\prime}}}|j^{\prime}\alpha^{\prime}\rangle$.
The second equation comes from the fact that for any $\mathcal{M}_m$,
there exists at most one nonzero term in each column which implies that
$\langle j\alpha|\mathcal{M}_m^{\dag}|j^{\prime}\alpha^{\prime}\rangle\langle k^{\prime}\beta^{\prime}|\mathcal{M}_m|j\alpha\rangle\neq0$
only if $j^{\prime}\alpha^{\prime}=k^{\prime}\beta^{\prime}$,
and the last equation comes from the relation $\sum\limits_m\mathcal{M}_m^{\dag}\mathcal{M}_m=\frac{|A^{\prime}|}{|A|}\mathbb{I}_{AB}$.
Then each diagonal element of the $J_{\phi'}:=J_{\frac{|A||B^{\prime}|}{|A^{\prime}||B|}\Theta^{\dag}\left(\Phi\right)}$ is
\begin{equation*}
\sum\limits_{m}\frac{|A||B^{\prime}|}{|A^{\prime}||B|}\langle j\alpha|\mathcal{M}_m^{\dag}(J_{\Phi})\mathcal{M}_m|j\alpha\rangle=\frac{1}{|B|}.
\end{equation*}
Thus we obtain $\Delta^{AB}(|B|J_{\phi'})=\mathbb{I}_{AB}$, i.e., for any $\Theta\in\mathcal{ISC}_{ABA^{\prime}B^{\prime}}$, we can always find a channel $\phi^{\prime}\in\mathcal{C}_{AB}$ such that $J_{\phi^{\prime}}\geqslant0, \Delta(|B|J_{\phi^{\prime}})=\mathbb{I}_{AB}$ corresponding to it.

On the other hand, for any $\phi'\in\mathcal{C}_{AB}$, consider the spectral decomposition of
\begin{eqnarray}\label{pfth11}
\frac{J_{\phi'}}{|A|}=\sum\limits_{j=1}^{|A|}\sum\limits_{\alpha=1}^{|B|}\lambda_{j,\alpha}|\psi_{j,\alpha}\rangle\langle\psi_{j,\alpha}|,
\end{eqnarray}
with $\sum\limits_{j=1}^{|A|}\sum\limits_{\alpha=1}^{|B|}\lambda_{j,\alpha}=1$, and $\lambda_{j,\alpha}\geqslant0$ for all $j$ and $\alpha$.
For each $|\psi_{j,\alpha}\rangle$, it can be written as
\begin{eqnarray}\label{pfth12}
|\psi_{j,\alpha}\rangle=\sum\limits_{k=1}^{|A|}\sum\limits_{\beta=1}^{|B|}c_{k,\beta}^{j,\alpha}|k\beta\rangle,
\end{eqnarray}
with $\sum\limits_{k=1}^{|A|}\sum\limits_{\beta=1}^{|B|}|c_{k,\beta}^{j,\alpha}|^2=1$.
For any  maximally coherent channel $\Phi\in\mathcal{C}_{A'B'}$, there exists $\{\theta_{j^{\prime}\alpha^{\prime}}\}_{j^{\prime}\alpha^{\prime}}$
such that $\frac{J_{\Phi}}{|A^{\prime}|}=|\Phi\rangle\langle\Phi|$,
with $|\Phi\rangle=\frac{1}{\sqrt{|A^{\prime}||B^{\prime}|}}\sum\limits_{j^{\prime}=1}^{|A^{\prime}|}
\sum\limits_{\alpha^{\prime}=1}^{|B^{\prime}|}e^{i\theta_{j^{\prime},\alpha^{\prime}}}|j^{\prime}\alpha^{\prime}\rangle$.
We define a map $\mathcal{K}_m^{j\alpha\dag}(\cdot)\mathcal{K}_m^{j\alpha}$ from $\mathcal{D}_{A'B'}$ to $\mathcal{D}_{AB}$ as
\begin{eqnarray}\label{pfth13}
\mathcal{K}_m^{j\alpha}=\sum\limits_{k=1}^{|A|}\sum\limits_{\beta=1}^{|B|}\mathcal{K}_{mk\beta}^{j\alpha}|f(k\beta)\rangle\langle k\beta|,~~m=1,\cdots,M,
\end{eqnarray}
where $\mathcal{K}_{mk\beta}^{j\alpha}=\sqrt{\frac{|A^{\prime}||B|}{M}}c_{k,\beta}^{j,\alpha}e^{i\theta_{f(k\beta)}}$ and $\langle f(j\alpha)|f(k\beta)\rangle=\delta_{jk,\alpha \beta}$.
Then we get
\begin{eqnarray*}
\frac{|A||B^{\prime}|}{|A^{\prime}||B|}\mathcal{K}_m^{j\alpha\dag}\left(J_{\Phi}\right)\mathcal{K}_m^{j\alpha}=\frac{J_{\psi_{j,\alpha}}}{M},
\end{eqnarray*}
where $J_{\psi_{j,\alpha}}=|A||\psi_{j,\alpha}\rangle\langle\psi_{j,\alpha}|$ is a desity operator of $\mathcal{D}_{AB}$.
Here, we should note that we can always find $f$ such that $\langle f(j\alpha)|f(k\beta)\rangle=\delta_{jk,\alpha \beta}$
holds since $|A||B|\leqslant|A^{\prime}||B^{\prime}|$.
Then we define the set of Kraus operators of the ISC $\Theta$ by $\{\mathcal{M}_{mj\alpha}|\mathcal{M}_{mj\alpha}:=\sqrt{\lambda_{j,\alpha}}\mathcal{K}_m^{j\alpha}\}$. We check that this is a well-defined set of Kraus operators of the incoherent superchannel $\Theta$ as follows
\begin{equation*}
\begin{array}{rl}
&\sum\limits_{mj\alpha}\mathcal{M}_{mj\alpha}^{\dag}\mathcal{M}_{mj\alpha}\\
& \ \ =\sum\limits_{mj\alpha}\sum\limits_{k\beta l\gamma}\lambda_{j\alpha}
\mathcal{K}_{mk\beta}^{j\alpha*}\mathcal{K}_{ml\gamma}^{j\alpha}|k\beta\rangle\langle f(k\beta)|f(l\gamma)\rangle\langle l\gamma|\\
& \ \ =\sum\limits_{mj\alpha}\sum\limits_{k\beta}\lambda_{j\alpha}|\mathcal{K}_{mk\beta}^{j\alpha}|^2|k\beta\rangle\langle k\beta|\\
& \ \ =\sum\limits_{j\alpha}\sum\limits_{k\beta}|A^{\prime}||B|\lambda_{j\alpha}|c_{k,\beta}^{j,\alpha}|^2|k\beta\rangle\langle k\beta|\\
& \ \ =\sum\limits_{j\alpha}\lambda_{j\alpha}|c_{k,\beta}^{j,\alpha}|^2\sum\limits_{k\beta}|A^{\prime}||B||k\beta\rangle\langle k\beta|\\
& \ \ =\sum\limits_{k\beta}|A^{\prime}||B|\frac{1}{|A||B|}|k\beta\rangle\langle k\beta|\\
& \ \ =\frac{|A^{\prime}|}{|A|}\mathbb{I}_{AB},
\end{array}
\end{equation*}
where the first equality is due to $\mathcal{M}_{mj\alpha}:=\sqrt{\lambda_{j,\alpha}}\mathcal{K}_m^{j\alpha}$ and Eq. (\ref{pfth13}). Combining the Eq. (\ref{pfth11}) and Eq. (\ref{pfth12}), and using the trace of $\frac{J_{\phi}}{|A|}$ is equal to 1, we get the fifth equality.
Thus, a superchannel $\Theta$ with its Kraus operators $\{\mathcal{M}_{mj\alpha}\}$ is an incoherent superchannel.
Moreover, 
$\frac{|A||B^{\prime}|}{|A^{\prime}||B|}\sum\limits_{mj\alpha}\mathcal{M}_{mj\alpha}^{\dag}\left(J_\Phi\right) \mathcal{M}_{mj\alpha}=\sum\limits_{j\alpha}\lambda_{j,\alpha} J_{\psi_{j,\alpha}}=J_{\phi'}$, i.e., $\frac{|A||B^{\prime}|}{|A^{\prime}||B|}\Theta^{\dag}\left(\Phi\right)=\phi'$,
with $\Delta^{AB}(|B|J_{\phi'})=\mathbb{I}_{AB}$.
Thus, for any operator $\phi'\in\mathcal{C}_{AB} $ with $J_{\phi'}\geqslant0$ and $\Delta^{AB}(|B|J_{\phi'})=\mathbb{I}_{AB}$, 
there exists an ISC $\Theta\in\mathcal{ISC}_{ABA^{\prime}B^{\prime}}$ such that
$\phi'=\frac{|A||B^{\prime}|}{|A^{\prime}||B|}\Theta^{\dag}\left(\Phi\right)$.

Thus, using Eq. (\ref{pfth}) and the one to one correspondence between $\{\Theta|\Theta\in\mathcal{ISC}_{ABA^{\prime}B^{\prime}}\}$ and $\{\phi^{\prime}|J_{\phi^{\prime}}\geqslant0, \Delta(|B|J_{\phi^{\prime}})=\mathbb{I}_{AB}, \phi^{\prime}\in\mathcal{C}_{AB}\}$, the right side of Eq. (\ref{max-4}) can be written as
\begin{eqnarray}\label{pfth21}
&& \ \ \ \frac{|B'|}{|A'|}\max\limits_{\Theta\in\mathcal{ISC}_{ABA^{\prime}B^{\prime}}}
F\left(\Theta(\phi),\Phi\right)^2\nonumber\\
&&=
\max\limits_{\substack{J_{\phi^{\prime}}\geqslant0,\\ \Delta(|B|J_{\phi'})=\mathbb{I}_{AB}}}
\tr\left[\frac{J_{\phi}}{|A|}(|B|J_{\phi'})\right].
\end{eqnarray}
The left side of Eq. (\ref{max-4}) can be written as
\begin{eqnarray}\label{pfth22}
2^{C_{\max}(\phi)}
&&=\min\limits_{\widetilde{\phi}\in\mathcal{IC}}\min\{\lambda:J_{\phi}\leqslant\lambda J_{\widetilde{\phi}}\}\nonumber\\
&&=\min\limits_{J_{\widetilde{\phi}}\geqslant0}
\left\{\tr\left[\frac{J_{\widetilde{\phi}}}{|A|}\right]|J_{\phi}\leqslant \Delta^{AB}(J_{\widetilde{\phi}}), J_{\widetilde{\phi}}\in M_{AB}
\right\}\nonumber\\
&&=\min\limits_{\substack{J_{\phi}\leqslant \Delta^{AB}(J_{\widetilde{\phi}})
\\ J_{\widetilde{\phi}}\in M_{AB},
J_{\widetilde{\phi}}\geqslant0}}
\tr\left[\frac{J_{\widetilde{\phi}}}{|A|}\right],
\end{eqnarray}
where $M_{AB}$ is the set of matrices that satisfies $\{M_{AB}:=kJ_\phi,~k>0,~\phi\in \mathcal{C}_{AB}\}$.

Our problem resuces to prove, for a given quantum channel $\phi\in\mathcal{C}_{AB}$,
\begin{equation}\label{pfth23}
\min\limits_{\substack{J_{\phi}\leqslant \Delta^{AB}(J_{\widetilde{\phi}})\\
J_{\widetilde{\phi}}\geqslant0,J_{\widetilde{\phi}}\in M_{AB}}}
\tr\left[\frac{J_{\widetilde{\phi}}}{|A|}\right]=
\max\limits_{\substack{J_{\phi^{\prime}}\geqslant0\\ \Delta^{AB}(|B|J_{\phi^{\prime}})=\mathbb{I}_{AB}}}
\tr\left[\frac{J_{\phi}}{|A|}(|B|J_{\phi^{\prime}})\right].
\end{equation}

Then combining Eq.(\ref{pfth21}) and Eq.(\ref{pfth22}), we complete the proof.

First, the right side of Eq. (\ref{pfth23}) can be written as
\begin{equation}\label{super-3}
\begin{array}{rl}
& \ \ \ \max\limits_{\substack{J_{\phi^{\prime}}\geqslant0\\ \Delta^{AB}(|B|J_{\phi^{\prime}})=\mathbb{I}_{AB}}}
\tr\left[\frac{J_{\phi}}{|A|}(|B|J_{\phi^{\prime}})\right]\\
&=
\max\limits_{\substack{J_{\phi^{\prime}}\geqslant0, J_{\phi^{\prime}}\in M_{AB}
\\ \Delta^{AB}(|B|J_{\phi^{\prime}})\leqslant\mathbb{I}_{AB}}}
\tr\left[\frac{J_{\phi}}{|A|}(|B|J_{\phi^{\prime}})\right].
\end{array}
\end{equation}
For $J_{\phi^{\prime}}\geqslant0$ with $\Delta^{AB}(|B|J_{\phi^{\prime}})\leqslant\mathbb{I}_{AB}$,
define $J_{\phi^{\prime\prime}}=J_{\phi^{\prime}}+\frac{\mathbb{I}_{AB}}{|B|}-\Delta^{AB}(J_{\phi^{\prime}})$,
then $J_{\phi^{\prime\prime}}\in M_{AB}$, $\Delta^{AB}(|B|J_{{\phi}^{\prime\prime}})=\mathbb{I}_{AB}$, and
$\tr\left[\frac{J_{\phi}}{|A|}(|B|J_{\phi^{\prime\prime}})\right]\geqslant\tr\left[\frac{J_{\phi}}{|A|}(|B|J_{\phi^{\prime}})\right]$.
Thus, Eq. \eqref{super-3} holds.
Now we  only need to prove
\begin{equation}\label{super-4}
\min\limits_{\substack{J_{\phi}\leqslant \Delta^{AB}(J_{\widetilde{\phi}})\\
J_{\widetilde{\phi}}\geqslant0, J_{\widetilde{\phi}}\in M_{AB}}}
\tr\left[\frac{J_{\widetilde{\phi}}}{|A|}\right]=
\max\limits_{\substack{J_{\phi^{\prime}}\geqslant0,J_{\phi^{\prime}}\in M_{AB}\\
\Delta^{AB}(|B|J_{\phi^{\prime}})\leqslant\mathbb{I}_{AB}}}
\tr\left[\frac{J_{\phi}}{|A|}(|B|J_{\phi^{\prime}})\right].
\end{equation}
The left side of Eq.\eqref{super-4} can be expressed as the following semidefinite programming
\begin{equation*}
\begin{array}{rl}
\min \tr\left[CJ_{\widetilde{\phi}}\right],\\[2.00mm]
s.t. \ \ \ \Lambda \left(J_{\widetilde{\phi}}\right)\geqslant D,\\[2.00mm]
J_{\widetilde{\phi}}\geqslant0,\\[2.00mm]
J_{\widetilde{\phi}}\in M_{AB},
\end{array}
\end{equation*}
where $C=\mathbb{I}_{AB}$, $D=J_{\phi}$, and $\Lambda=\Delta^{AB}$.
Then the dual semidefinite programming is given by
\begin{equation*}
\begin{array}{rl}
\max \tr\left[D\left(|B|J_{\phi^{\prime}}\right)\right],\\[2.00mm]
s.t. \ \ \ \Lambda^{\dag}\left(|B|J_{\phi^{\prime}}\right)\leqslant C,\\[2.00mm]
J_{\phi^{\prime}}\geqslant0\\[2.00mm]
J_{\phi^{\prime}}\in M_{AB}.
\end{array}
\end{equation*}
That is,
\begin{equation*}
\begin{array}{rl}
\max\tr\left[J_{\phi}\left(|B|J_{\phi^{\prime}}\right)\right],\\
s.t. \ \ \ \Delta^{AB}(|B|J_{\phi^{\prime}})\leqslant \mathbb{I}_{AB},\\
J_{\phi^{\prime}}\geqslant0\\
J_{\phi^{\prime}}\in M_{AB}.
\end{array}
\end{equation*}
Note that the dual is strictly feasible as we only need to choose
$J_{\widetilde{\phi}}=2\lambda_{\max}(J_{\phi})\mathbb{I}_{AB}\in M_{AB}$, where $\lambda_{\max}(J_{\phi})$ is the maximum eigenvalue of $J_{\phi}$.
Thus, strong duality holds, and the Eq. ({\ref{pfth23}}) and Eq. (\ref{super-4}) are proved.

\section*{D: Proof of that Theorem 1 holds for the set of $\mathcal{DISC}$ and $\mathcal{SISC}$}

In this section, we will prove Theorem 1 also established for the set of $\mathcal{DISC}$ and $\mathcal{SISC}$. From the derivation of Theorem 1, we just need to get the one to one corresponding relation for the set of $\mathcal{DISC}_{ABA^{\prime}B^{\prime}}$ and the set
$\{\phi^{\prime}|J_{\phi^{\prime}}\geqslant0, \Delta(|B|J_{\phi^{\prime}})=\mathbb{I}_{AB}\}$. So as the $\mathcal{SISC}$ and $\{\phi^{\prime}|J_{\phi^{\prime}}\geqslant0, \Delta(|B|J_{\phi^{\prime}})=\mathbb{I}_{AB}\}$.

For any  $\Theta\in\mathcal{DISC}_{ABA^{\prime}B^{\prime}}$, from the definition of $\Theta\in\mathcal{DISC}_{ABA^{\prime}B^{\prime}}$, one has $\Delta^{A^{\prime}B^{\prime}}\Theta=\Theta\Delta^{AB}$,
then one can get that $\Delta^{AB}\Theta^{\dag}=\Theta^{\dag}\Delta^{A^{\prime}B^{\prime}}$,
which implies $\Delta^{AB}\left(\sum\limits_m\mathcal{M}_m^{\dag}(\cdot)\mathcal{M}_m\right)=
\sum\limits_m\mathcal{M}_m^{\dag}\left(\Delta^{A^{\prime}B^{\prime}}(\cdot)\right)\mathcal{M}_m$,
with $\Theta=\{\mathcal{M}_m\}_m$.
Thus, we have
\begin{equation*}
\begin{array}{rl}
\Delta^{AB}\left(|B|J_{\phi'}\right)&=|B|\frac{|A||B^{\prime}|}{|A^{\prime}||B|}\Delta^{AB}
\left(\sum\limits_m\mathcal{M}_m^{\dag}J_{\Phi}\mathcal{M}_m\right)\\
&=\frac{|A||B^{\prime}|}{|A^{\prime}|}\left(\sum\limits_m\mathcal{M}_m^{\dag}\left(\Delta^{A^{\prime}B^{\prime}}(J_{\Phi})\right)\mathcal{M}_m\right)\\
&=\frac{|A||B^{\prime}|}{|A^{\prime}|}\sum\limits_m\mathcal{M}_m^{\dag}
\left(\frac{1}{|B^{\prime}|}\mathbb{I}_{A^{\prime}B^{\prime}}\right)\mathcal{M}_m\\
&=\mathbb{I}_{AB},
\end{array}
\end{equation*}
where the second equation obtained by using the property $\Delta^{AB}\Theta^{\dag}=\Theta^{\dag}\Delta^{A^{\prime}B^{\prime}}$.
The last equation comes from the relation
$\sum\limits_m\mathcal{M}_m^{\dag}\mathcal{M}_m=\frac{|A^{\prime}|}{|A|}\mathbb{I}_{AB}$.

Similar to the proof for the part of Theorem 1: for any operator $\phi'\in\mathcal{C}_{AB} $ with $J_{\phi'}\geqslant0$ and $\Delta^{AB}(|B|J_{\phi'})=\mathbb{I}_{AB}$, 
there exists an ISC $\Theta\in\mathcal{ISC}_{ABA^{\prime}B^{\prime}}$ such that
$\phi'=\frac{|A||B^{\prime}|}{|A^{\prime}||B|}\Theta^{\dag}\left(\Phi\right)$.
Here we just change the expression of Eq. (\ref{pfth13}) $\mathcal{K}_m^{j\alpha}$ as
$\mathcal{K}_m^{j\alpha}=\sum\limits_{k=1}^{|A|}\sum\limits_{\beta=1}^{|B|}\mathcal{K}_{mk\beta}^{j\alpha}|k\beta\rangle\langle k\beta|$
with $\mathcal{K}_{mk\beta}^{j\alpha}=\sqrt{\frac{|A^{\prime}||B|}{M}}c_{k,\beta}^{j,\alpha}$. One can get the desired result:  for any operator $\phi'\in\mathcal{C}_{AB} $ with $J_{\phi'}\geqslant0$ and $\Delta^{AB}(|B|J_{\phi'})=\mathbb{I}_{AB}$, 
there exists an DISC $\Theta\in\mathcal{DISC}_{ABA^{\prime}B^{\prime}}$ such that
$\phi'=\frac{|A||B^{\prime}|}{|A^{\prime}||B|}\Theta^{\dag}\left(\Phi\right)$.
As for $\mathcal{SISC}$, similar to the proof of incoherent superchannels $\mathcal{SIC}$, one can also obtain the one to one corresponding relation between the set of $\mathcal{SISC}_{ABA^{\prime}B^{\prime}}$ and the set
$\{\phi^{\prime}|J_{\phi^{\prime}}\geqslant0, \Delta(|B|J_{\phi^{\prime}})=\mathbb{I}_{AB}\}$.
Therefore, Theorem 1 can also be established for the set of $\mathcal{DISC}$ and $\mathcal{SISC}$.

\section*{E: Proof of Theorem 2.}
{\it Restatement of Theorem 2.} --For any quantum channel $\phi\in\mathcal{C}_{AB}$, $2^{C_{\max}(\phi)}$ is the maximal advantage achievable by $\phi$
compared with all incoherent channels in sub-superchannel discriminations of incoherent superchannel instruments:
\begin{equation*}
2^{C_{\max}(\phi)}=\max\limits_{\mathfrak{J}}\frac{p_{succ}(\mathfrak{J},\phi)}{p_{succ}^{ISCO}(\mathfrak{J})},
\end{equation*}
where $\mathfrak{J}$ denotes the incoherent instrument of superchannel $\Theta\in\mathcal{ISC}_{ABA^{\prime}B^{\prime}}$, $|A||B|\leqslant|A^{\prime}||B^{\prime}|$.

{\sf Proof.} Due to the definition of $C_{\max}(\phi)$, there exists an incoherent channel $\widetilde{\phi}$ such that
$J_{\phi}\leqslant 2^{C_{\max}(\phi)}J_{\widetilde{\phi}}$.
Thus, for any incoherent instrument of superchannel $\mathfrak{J}$ and POVM
$\{M_{k}\}_{k}$ on $\phi,\widetilde{\phi}\in\mathcal{C}_{AB}$,
\begin{equation*}
p_{succ}(\mathfrak{J},\{M_{k}\}_{k},J_{\phi})\leqslant
2^{C_{\max}(\phi)}p_{succ}(\mathfrak{J},\{M_{k}\}_{k},J_{\widetilde{\phi}}).
\end{equation*}
Then
\begin{equation*}
p_{succ}(\mathfrak{J},J_{\phi})\leqslant2^{C_{\max}(\phi)}p_{succ}^{ISCO}(\mathfrak{J}),
\end{equation*}
where $p_{succ}^{ISCO}(\mathfrak{J})$ the optimal probability of success among all incoherent channels.

Our task now is to acheve upper bound for discrimination.
From Theorem 1, there exists an incoherent superchannel
$\Theta=\{\mathcal{M}_m\}_m\in\mathcal{ISC}_{ABA^{\prime}B^{\prime}}$ such that
\begin{eqnarray}\label{pf2}
2^{C_{\max}(\phi)}=\frac{|B'|}{|A'|}F\left(\Theta(\phi),\Phi\right)^2,
\end{eqnarray}
with $\Phi$ a maximally coherent channel.
Consider unitaries $\{\mathcal{U}_k\},~k=1,2,\ldots,|A^{\prime}||B^{\prime}|$, with
\begin{equation*}
\mathcal{U}_k=\sum\limits_{j^{\prime}=1}^{|A^{\prime}|}\sum\limits_{\alpha^{\prime}=1}^{|B^{\prime}|}
e^{i\frac{j^{\prime}\alpha^{\prime}}{|A^{\prime}||B^{\prime}|}2k\pi}|j^{\prime}\alpha^{\prime}\rangle\langle j^{\prime}\alpha^{\prime}|.
\end{equation*}
Define the sub-superchannel $\Theta_k=\{\mathcal{M}_{mk}\}_m$ with
$\mathcal{M}_{mk}=\frac{1}{\sqrt{|A^{\prime}||B^{\prime}|}}\mathcal{U}_k\mathcal{M}_m$.
Then we have
\begin{equation*}
J_{\Theta_k(\phi)}=\frac{1}{|A^{\prime}||B^{\prime}|}\mathcal{U}_kJ_{\Theta(\phi)}\mathcal{U}_k^{\dag}.
\end{equation*}
Thus, the superchannel $\widetilde{\Theta}=\sum\limits_{k}\Theta_k$ is also an $\mathrm{ISC}$.
That is, the instrument $\widetilde{\mathfrak{J}}=\{\Theta_k\}_k$ is an incoherent instrument of $\widetilde{\Theta}$.

For any POVM $\{M_k\}_{k}$ and any incoherent channel $\widetilde{\phi}$,
one has
\begin{equation*}
\begin{array}{rl}
&p_{succ}(\widetilde{\mathfrak{J}},\{M_{k}\}_{k},\widetilde{\phi})\\[3.00mm]
& \ =\frac{1}{|A^{\prime}|}\sum\limits_{k}\tr
\left[\left(\sum\limits_m\mathcal{M}_{mk}J_{\widetilde{\phi}}\mathcal{M}_{mk}^{\dag}\right)M_{k}\right]\\[3.0mm]
& \ =\frac{1}{|A^{\prime}|^2|B^{\prime}|}\tr\left[\left(\sum\limits_m\mathcal{M}_mJ_{\widetilde{\phi}}M_{m}^{\dag}\right)
\left(\sum\limits_{k}\mathcal{U}_k^{\dag}M_{k}\mathcal{U}_k\right)\right]\\[3.00mm]
& \ =\frac{1}{|A^{\prime}|^2|B^{\prime}|}\tr\left[\left(\sum\limits_m\mathcal{M}_m J_{\widetilde{\phi}}\mathcal{M}_m^{\dag}\right)
\left(\sum\limits_{k}\mathcal{U}_k^{\dag}\Delta(M_{k})\mathcal{U}_k\right)\right]\\[3.00mm]
& \ =\frac{1}{|A^{\prime}|^2|B^{\prime}|}\tr\left[\left(\sum\limits_m\mathcal{M}_m J_{\widetilde{\phi}}\mathcal{M}_m^{\dag}\right)
\sum\limits_{k}\Delta(M_{k})\right]\\[3.00mm]
& \ =\frac{1}{|A^{\prime}|^2|B^{\prime}|}\tr\left[\left(\sum\limits_m\mathcal{M}_m J_{\widetilde{\phi}}\mathcal{M}_m^{\dag}\right)
\Delta(\sum\limits_{k}M_{k})\right]\\[3.00mm]
& \ =\frac{1}{|A^{\prime}|^2|B^{\prime}|}\tr\left[\sum\limits_m\mathcal{M}_m J_{\widetilde{\phi}}\mathcal{M}_m^{\dag}\right]\\[3.00mm]
& \ =\frac{1}{|A^{\prime}||B^{\prime}|}.
\end{array}
\end{equation*}
That is to say, $p_{succ}^{ISCO}(\widetilde{\mathfrak{J}})=\frac{1}{|A^{\prime}||B^{\prime}|}$.
Taking another POVM $\{N_k\}_k$ on $\phi\in\mathcal{C}_{AB }$ as
\begin{equation*}
N_k=\mathcal{U}_k|\Phi\rangle\langle\Phi|\mathcal{U}_k^{\dag},
\end{equation*}
where $J_\Phi=|A'||\Phi\rangle\langle\Phi|$ is the Choi matrix of the maximally coherent channel in Eq. (\ref{pf2}).

Then we have
\begin{equation*}
\begin{array}{rl}
&p_{succ}(\widetilde{\mathfrak{J}},\{N_{k}\}_k,\phi)\\[3.00mm]
& \ =\frac{1}{|A^{\prime}|}\sum\limits_k\tr\left[\left(\sum\limits_m\mathcal{M}_{mk}J_{\phi}\mathcal{M}_{mk}^{\dag}\right)N_k\right]\\[3.00mm]
& \ =\frac{1}{|A^{\prime}|^2|B^{\prime}|}\tr\left[\left(\sum\limits_m\mathcal{M}_{m}J_{\phi}\mathcal{M}_{m}^{\dag}\right)
\left(\sum\limits_k\mathcal{U}_k^{\dag}N_k\mathcal{U}_k\right)\right]\\[3.00mm]
& \ =\frac{1}{|A^{\prime}|^2|B^{\prime}|}\tr\left[\left(\sum\limits_m\mathcal{M}_{m}J_{\phi}\mathcal{M}_{m}^{\dag}\right)
\sum\limits_k|\Phi\rangle\langle\Phi|\right]\\[3.00mm]
& \ =\frac{1}{|A^{\prime}|}\tr\left[J_{\Theta(\phi)}|\Phi\rangle\langle\Phi|\right]\\[2.00mm]
& \ =\frac{1}{|A^{\prime}|^2}\tr\left[J_{\Theta(\phi)}J_\Phi\right]\\
& \ =\frac{1}{|A^{\prime}||B^{\prime}|}\frac{|B^{\prime}|}{|A^{\prime}|}\tr\left[J_{\Theta(\phi)}J_\Phi\right]\\
& \ =\frac{1}{|A^{\prime}||B^{\prime}|}\frac{|B'|}{|A'|}F\left(\Theta(\phi),\Phi\right)^2\\
& \ =\frac{2^{C_{\max}(\phi)}}{|A^{\prime}||B^{\prime}|}\\[2.00mm]
& \ =2^{C_{\max}(\phi)}p_{succ}^{ISCO}(\widetilde{\mathfrak{J}}).
\end{array}
\end{equation*}
 Thus, for this incoherent instrument $\widetilde{\mathfrak{J}}=\{\Theta_k\}_k$, $p_{succ}(\widetilde{\mathfrak{J}},\phi)\geq 2^{C_{\max}(\phi)}p_{succ}(\widetilde{\mathfrak{J}})$.
We have thus proved the Theorem 2.
With a similar consideration, one can see the result holds for $\mathcal{DISC}_{ABA^{\prime}B^{\prime}}$ and $\mathcal{SISC}_{ABA^{\prime}B^{\prime}}$.

\section*{F: Proof of properties of $C_R$}
\subsection*{I: Proof of conditions (iii) and (iv)}
We define the robustness of coherence (ROC) of a quantum channel $\phi\in\mathcal{C}_{AB}$ as
\begin{equation}\label{roc}
C_R(\phi)=\min_{\widetilde{\phi}\in\mathcal{C}_{AB}}\Big\{s\geq0\Big|\frac{J_{\phi}+sJ_{\widetilde\phi}}{1+s}=J_{\phi^\prime}, \phi^\prime\in \mathcal{IC}_{AB}\Big\},
\end{equation}
that is, the minimum weight of another channel $\widetilde{\phi}$ such that its convex mixture with $\phi$ yields an incoherent channel $\phi^\prime$. 
We now prove that the ROC is a bona fide measure of coherence for channels. First of all, it is seen by definition that
$C_R(\phi)\geq0$ and $C_R(\phi)=0$ if and only if $\phi\in\mathcal{IC}_{AB}$.

Second, the ROC is convex. Let $\phi_1$ and $\phi_2$ be two channels, and write for each the optimal pseudomixture
\begin{eqnarray*}
J_{\phi_k}=[1+C_R(\phi_k)]J_{\phi_k^{\prime*}}-C_R(\phi_k)J_{\widetilde\phi^*_k},
\end{eqnarray*}
with $k=1,2$.
Taking the convex combination $\phi=p\phi_1+(1-p)\phi_2$ with $0\leq p\leq 1$, notice that
\begin{eqnarray*}
J_\phi=(1+s)J_{\phi^\prime}-sJ_{\widetilde\phi}
\end{eqnarray*}
can be written with
\begin{eqnarray*}
J_{\phi^\prime}=\frac{p[(1+C_R(\phi_1)]J_{\phi_1^{\prime*}}+(1-p)[(1+C_R(\phi_2)]J_{\phi_2^{\prime*}}}{1+s},
\end{eqnarray*}
then $\phi^\prime\in \mathcal{IC}_{AB}$, $J_{\widetilde\phi}=[pC_R(\phi_1)J_{\widetilde\phi^*_1}+(1-p)C_R(\phi_2)J_{\widetilde\phi^*_2}]/s$ and $s=pC_R(\phi_1)+(1-p)C_R(\phi_2)$. By definition, $C_R(\phi)\leq s$, which proves convexity,
\begin{eqnarray*}
C_R(p\phi_1+(1-p)\phi_2)\leq pC_R(\phi_1)+(1-p)C_R(\phi_2).
\end{eqnarray*}

Third, the ROC is nonincreasing on average under incoherent superchannels. Let  $\Theta=\{\mathcal{M}_m\}_m\in\mathrm{ISC}_{ABA^{\prime}B^{\prime}}$. For any channel $\phi$, let $\widetilde\phi^*$ and $\phi^{\prime*}$ denote the channels in the optimal pseudomixture for $C_R(\phi)$ as in Eq. (\ref{roc}), and apply $\mathcal{M}_m$ on both sides, we get
\begin{eqnarray*}
&&\mathcal{M}_mJ_{\phi}\mathcal{M}_m^{\dag}\nonumber\\
&&=[1+C_R(\phi)]\mathcal{M}_mJ_{\phi^{\prime*}}\mathcal{M}_m^{\dag}-C_R(\phi)\mathcal{M}_mJ_{\widetilde\phi^*}\mathcal{M}_m^{\dag}.
\end{eqnarray*}
By defining
\begin{eqnarray*}
J_{\phi_m^{\prime*}}=\frac{1}{1+s_m}\frac{1}{p_m}[1+C_R(\phi)]\mathcal{M}_mJ_{\phi^{\prime*}}\mathcal{M}_m^{\dag},
\end{eqnarray*}
\begin{eqnarray*}
J_{\widetilde\phi^*_m}=\frac{1}{s_m}\frac{1}{p_m}C_R(\phi)\mathcal{M}_mJ_{\widetilde\phi^*}\mathcal{M}_m^{\dag},
\end{eqnarray*}

\begin{eqnarray*}
s_m=\frac{1}{p_m}C_R(\phi)\frac{\mathrm{Tr}(\mathcal{M}_mJ_{\widetilde\phi^*}\mathcal{M}_m^{\dag})}{|A'|},
\end{eqnarray*}
with $p_m=\frac{\mathrm{Tr}(\mathcal{M}_mJ_{\phi}\mathcal{M}_m^{\dag})}{|A'|}$, we can write
\begin{eqnarray*}
J_{\phi_m}=(1+s_m)J_{\phi_m^{\prime*}}-s_mJ_{\widetilde\phi_m^*},
\end{eqnarray*}
where $J_{\phi_m}=\mathcal{M}_mJ_{\phi}\mathcal{M}_m^{\dag}/p_m$. Since the latter pseudomixture for each $\phi_m$ is not necessarily optimal, it follows by Eq. (\ref{roc}) that $C_R(\phi_m)\leq s_m=\frac{1}{p_m}C_R(\phi)\frac{\mathrm{Tr}(\mathcal{M}_mJ_{\widetilde\phi^*}\mathcal{M}_m^{\dag})}{|A'|}$.
Taking the weighted average over all sub-superchannels, we finally get
\begin{eqnarray*}
&&\sum_mp_mC_R(\phi_m)\nonumber\\
&&\leq\sum_mp_m\frac{1}{p_m}C_R(\phi)\frac{\mathrm{Tr}(\mathcal{M}_mJ_{\widetilde\phi^*}\mathcal{M}_m^{\dag})}{|A'|},\nonumber\\
&&=C_R(\phi).
\end{eqnarray*}

\subsection*{II: Derivation of (8)}
In the following, we give ROC of a channel $C_R(\phi)$ another expression as
\begin{eqnarray}\label{lroc}
C_R(\phi)=\min_{\phi^\prime\in \mathcal{IC}_{AB}}\{s\geq0|J_\phi\leq (1+s)J_{\phi^\prime}\}.
\end{eqnarray}
From the definition of ROC of a channel, one has
\begin{eqnarray*}\label{opt}
J_{\phi}=[1+C_R(\phi)]J_{\phi^{\prime*}}-C_R(\phi)J_{\widetilde\phi^*},
\end{eqnarray*}
with $\widetilde\phi^*$ and $\phi^{\prime*}$ the channels in the optimal pseudomixture for $C_R(\phi)$ as in Eq. (\ref{roc}). This implies
\begin{eqnarray*}
J_{\phi}\leq[1+C_R(\phi)]J_{\phi^{\prime*}},
\end{eqnarray*}
with $\phi^{\prime*}\in \mathcal{IC}_{AB}$, which means that $C_R(\phi)$ is lower-bounded by the minimum on the right-hand side of Eq. (\ref{lroc}). On the other hand, suppose $J_\phi\leq (1+s)J_{\phi^{\prime}}$ for some $\phi^{\prime}\in \mathcal{IC}_{AB}$. Then we can write
\begin{eqnarray*}
J_{\phi^{\prime}}=\frac{J_{\phi}+sJ_{\widetilde\phi}}{1+s},
\end{eqnarray*}
with $J_{\widetilde\phi}=[(1+s)J_{\phi^{\prime}}-J_\phi]/s$ a valid Choi matrix of a channel $\widetilde\phi$. This proves that the minimum in Eq. (\ref{lroc}) is also an upper bound for $C_R(\phi)$, hence we conclude that (\ref{lroc}) holds.